# MAGNETIC FIELD OF SUPERCONDUCTIVE IN-VACUO UNDULATORS IN COMPARISON WITH PERMANENT MAGNET UNDULATORS

Herbert O. Moser[a], Robert Rossmanith[a,b*]

[a]National University of Singapore, Singapore Synchrotron Light Source, Singapore

[b]Synchrotron Radiation Research Group, Forschungszentrum Karlsruhe, P. O. Box 3640, D-76021 Karlsruhe, Germany

## Abstract

During the last few years superconductive undulators with a period length of 3.8 mm and 14 mm have been built. In this paper scaling laws for these novel insertion devices are presented: a simple analytic formula is derived which describes the achievable magnetic field of a superconcuctive undulator as a function of gap-width and period length.

*Corresponding author. Tel.: +49-7247-82-6179, fax: +49-7247-82-6172

E-mail address: rossmanith@anka.fzk.de (R. Rossmanith)

# Introduction

During the last few years institutes involved in the development of synchrotron light sources and FELs have shown increasing interest in superconductive in-vacuo undulators. Efforts to build such devices started at various labs around 1990 [1]. In Karlsruhe activities have focused on technical design studies for a 3.8 mm period length undulator [2] followed by beam tests at the Mainz Microtron MAMI [3] and the construction of two undulators each with a period length of 14 mm for the Singapore Synchrotron Light Source [4] and for ANKA [5]. These studies have resulted in the emergence of a clear concept of how to build such superconductive undulators and the cryostats for the various applications.

The aim of this paper is not to describe the technical layout but to present simple scaling laws showing how the maximum field strength varies with the period length, the gap width, the undulator current and the pole width. With these formulas the user of such devices can evaluate the range of obtainable fields and k-values similar to the formulas given for permanent magnet undulators in literature [6]. A comparison of permanent undulators and superconductive undulators from a different point of view can be found in [7].

# Summary on permanent magnet undulators

It is well known that the magnetic field of permanent magnet undulators (pure $SmCo_5$ undulators) can be described by the approximation [6]

$$B_{max} [T] = 1.55 \exp\left(\frac{-\pi g}{p}\right) \qquad (1)$$

Where $B_{max}$ is the maximum field in T, g is the width of the gap in mm and p is the period length in mm. Similar formulas are valid for hybrid undulators:

$$B_{max} [T] = 0.95 \, a \, \exp\left(\frac{-g}{p}\left(b - c\frac{g}{p}\right)\right) \qquad (2)$$

For $SmCo_5$ a = 3.33, b = 5.47, c = 1.8, for a Nd-Fe-B a = 3.44, b = 5.08 and c = 1.54.

## Superconductive Undulators

The model used for the field calculations for superconductive undulators is sketched in fig. 1 and 2. Only the section of the undulator close to the beam is shown. The darker parts are made of iron, the brighter parts are superconducting wires. The current direction in the superconductor is alternating. The different colours of the superconductor mark the different directions of the current. The matching sections are at the beginning and the end of the undulator. The parameters used in the following are defined in fig. 3.

All computations presented in this paper are done with the help of the program SRW developed by the ESRF [8].

The units for the current density I in the superconductor are $kA/mm^2$. For an undulator with a 14 mm period length the calculated maximum field is shown in fig. 4. In these calculations it was assumed that the pole width w is 2 mm and the current density varies.

The dependence of the maximum field on the current density is shown in fig. 5 (gap width g = 5 mm). The curve shows the two regions of the undulator: the region below circa 0.2 $kA/mm^2$ where the iron is not saturated and the region where the iron is saturated. For all practical purposes the region in which the iron is saturated is the most important one. In this region the field can be approximated by the function

$$B_{max} [T] = 0.023 + 0.045 \cdot I + (9.53 + 7.75 \cdot I) \exp(-0.51 \cdot g) \qquad (3)$$

I in kA/mm$^2$ and g in mm. This formula is valid for a 14 mm period length undulator with a 2 mm pole width w.

The maximum field as a function of pole width w and period length p is shown in fig. 6. The current density in fig. 6 is 1 kA/mm$^2$. The optimum pole width w is between 2.7 and 2.8 mm. The maximum of the curve is rather flat.

The final formula describing the behaviour of the superconductive undulator as a function of period length p, gap width g and current density I is:

$$B_{max}[T] = [(0.023 + 0.045 \cdot I) + (9.55 + 7.75 \cdot I) \cdot \exp(-0.51 \cdot g)] \cdot \left( \frac{-1.023 + 26.3 \cdot \exp(-0.8 \cdot g) + p \cdot (0.11 + 0.21 \exp(-0.43 \cdot g))}{0.517 + 26.3 \cdot \exp(-0.8 \cdot g) + 2.94 \cdot \exp(-0.43 \cdot g)} \right) \quad (4)$$

This formula is compared in fig. 7 with the formula for a permanent magnet undulator. The superconductive undulator has a significantly higher field over the permanent magnet undulator.

NbTi-wires have been used in all prototypes built up to now. The maximum current is limited by the magnetic field at the surface of the superconductor. Fig. 8 shows the calculated field distribution inside the coiling. The dark bars mark the position of the iron. The maximum field on the surface of the superconductor is about 2.2 Tesla. Commercial NbTi superconductors can operate at these fields with current densities of 1.3 to 1.4 kA [2]. Therefore, curve C in fig. 7 marks the upper limit for NbTi wires.

## Summary

The maximum magnetic field produced by superconductive undulators for a given gap and a given period length is higher than the field in permanent undulators. The relation between the parameters current density, gap width and period length is summarized in the analytic approximation in equation (4).

References

[1] I. Ben-Zvi et al., The performance of a superconductive micro-undulator prototype, Nucl. Instr. Meth. A297(1990)301

H. O. .Moser, B. Krevet, H. Holzapfel, Mikroundulator, German Patent P101 094.9-33 Jan. 16, 1991

[2] T. Hezel et al., A superconductive undulator with a period length of 3.8 mm, J. Synchrotron Rad. (1998) 5 448

[3] T. Hezel et al., Experimental results with a novel superconductive in-vacuum mini-undulator test device at the Mainz microtron MAMI, Proc. 1999 Particle Accelerator Conference, New York 1999

[4] H. O. Moser et al., Status and Planned Development of the Singapore Synchrotron Light Source, Asian Particle Accelerator Conference APAC'01, Sept. 2001

[5] H. O. Moser, R. Rossmanith, Design Study of a superconductive in-vacuo undulator for storage rings with an electrical tunability between 0 and 2, Proceedings European Accelerator Conference, EPAC 2000, Vienna June 2000

[6] J.Murphy, Synchrotron Light Source Data Book, Internal Report, BNL42333J

[7] P. Elleaume, J. Chavanne, Bart Faatz, Design Considerations for a 1 Angstroem SASE undulator, Nucl. Instr. Meth. A455 (2000)503

[8] P. Elleaume, O. Chubar and J. Chavanne, Computing 3D MagneticFields from Insertion Devices, Proceedings Particle Accelerator Conference, Vancouver, May 1997

Figure Captions

Fig. 1 Model of the superconductive undulator. Dark colors represent magnetic iron, light colors represent superconductive material.

Fig. 2: Different view of fig. 1. Matching sections are at the beginning and the end of the undulator.

Fig. 3: Definition of gap-width g, pole-width w and period length p

Fig. 4 Maximum field of a superconductive undulator with a period length p of 14 mm and a pole width w of 2 mm. Current density: A= 1.4 kA/mm$^2$, B= 1.2 kA/mm$^2$, C= 1. kA/mm$^2$, D= 0.8 kA/mm$^2$. E shows the maximum field of a permanent magnet undulator according to formula (1).

Fig. 5 Dependence of the maximum field on the current density I of the superconductive wire. Gap width g is 5 mm, period length p is 14 mm, pole width w is 2 mm

Fig. 6 Maximum magnetic field in Tesla as a function of period length p and pole width w, I =1 kA/mm$^2$, gap width g = 5 mm.

Fig. 7 Comparison of maximum field B[T] as a function of gap width g and period length p for a permanent magnet undulator A, a superconductive undulator with a current density of 1 kA/mm$^2$ B and for a superconductive undulator with the current density of 1.4 kA/mm$^2$ C

Fig. 8 The field inside the coiling determines the maximum current density I = 1.4 kA. The dark bars mark the iron poles. The field increases towards the poles and has its maximum in the iron. The boundary between the poles and superconductive wires determines the maximum current.

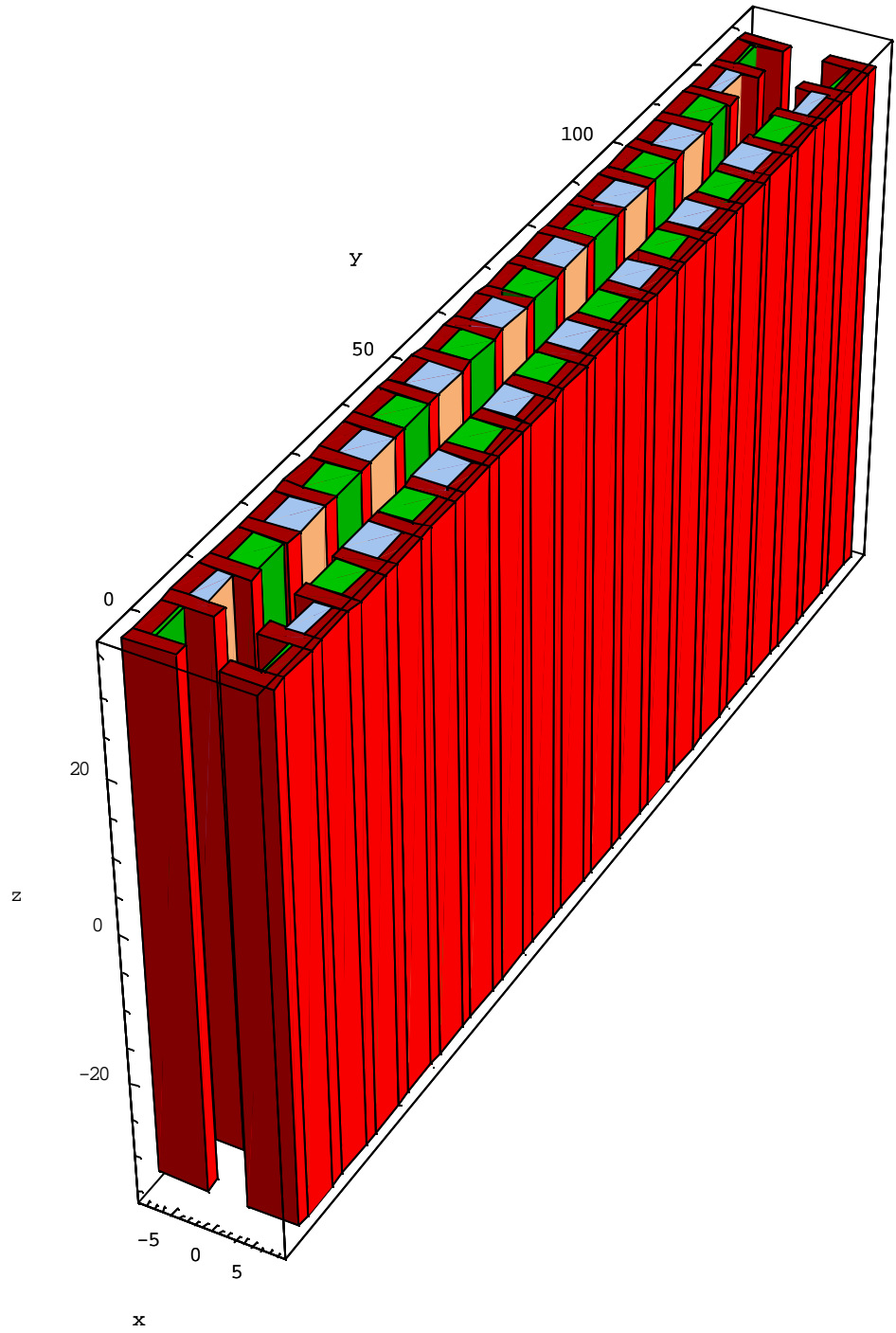

Fig. 1

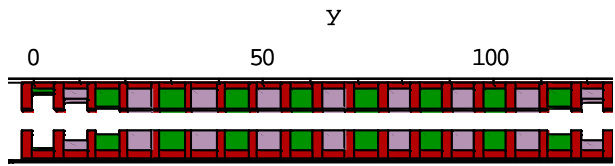

Fig. 2

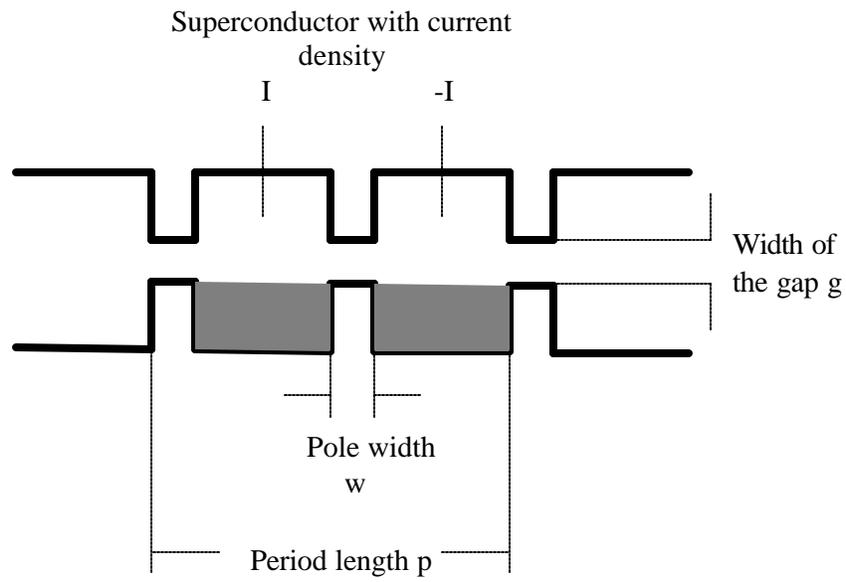

Fig. 3

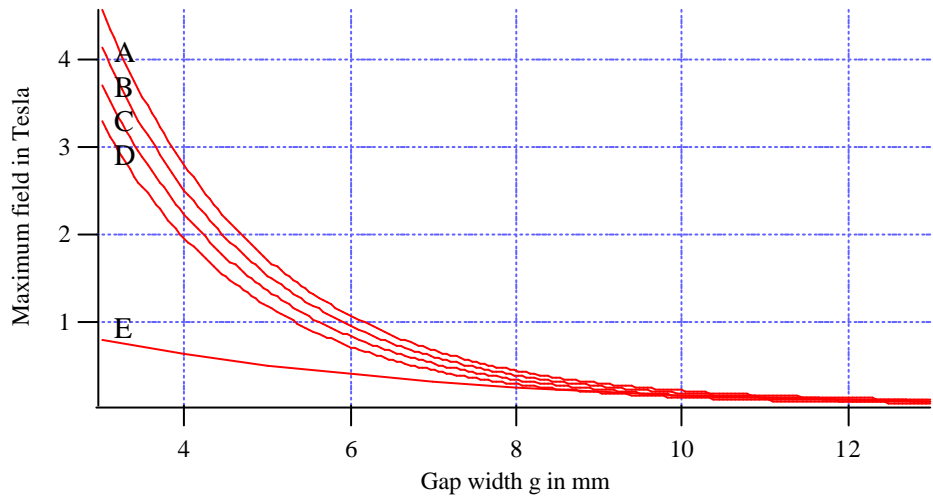

Fig. 4

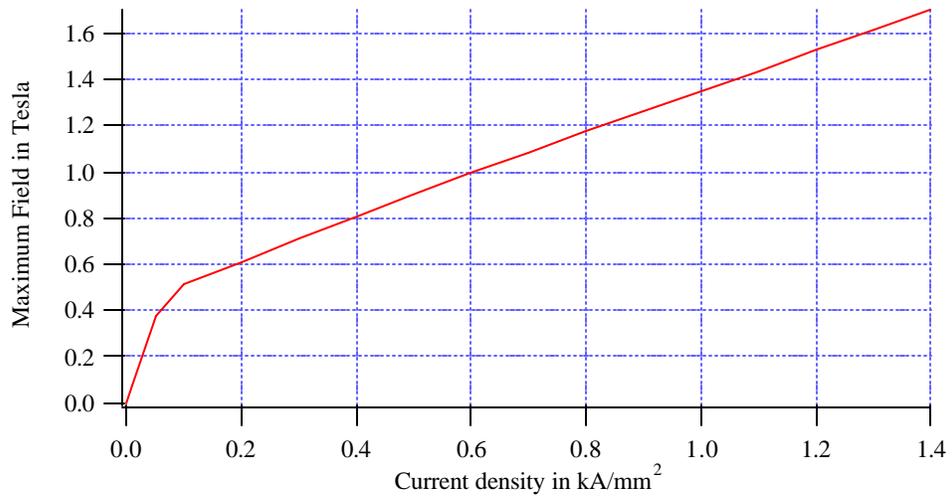

Fig. 5

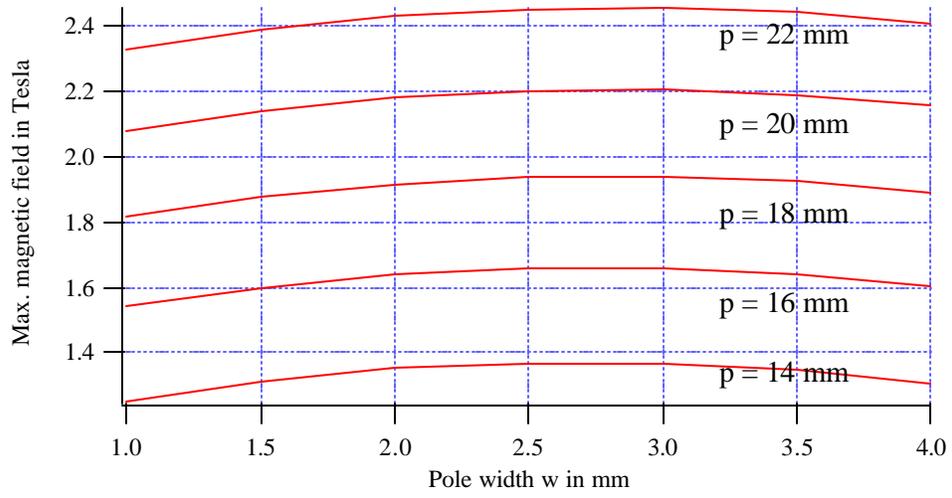

Fig. 6

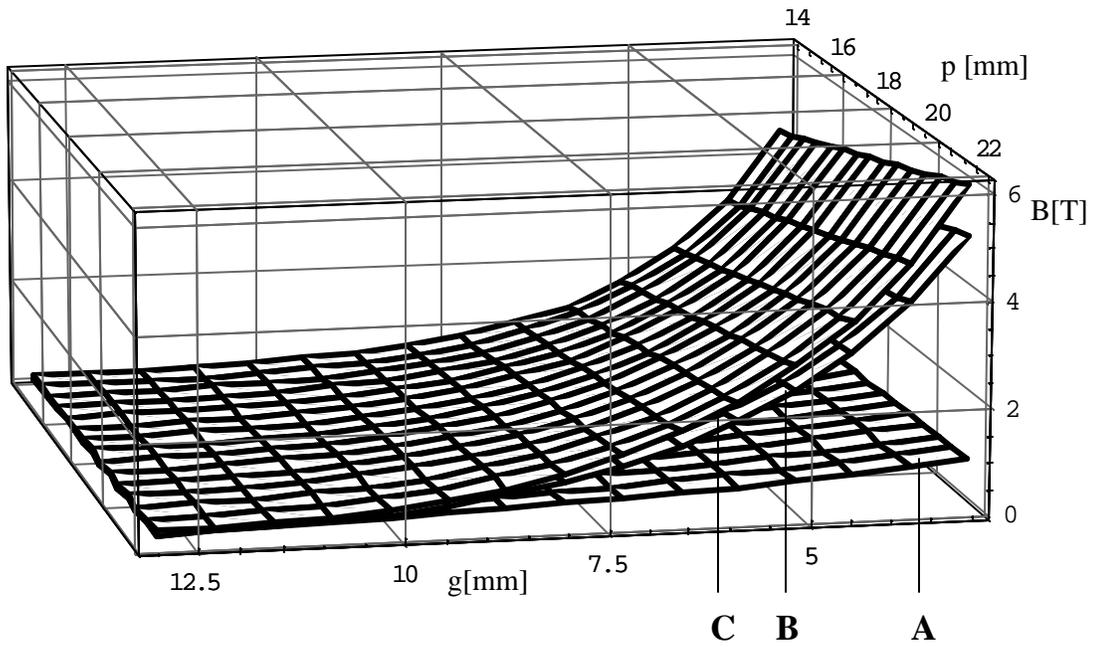

Fig. 7

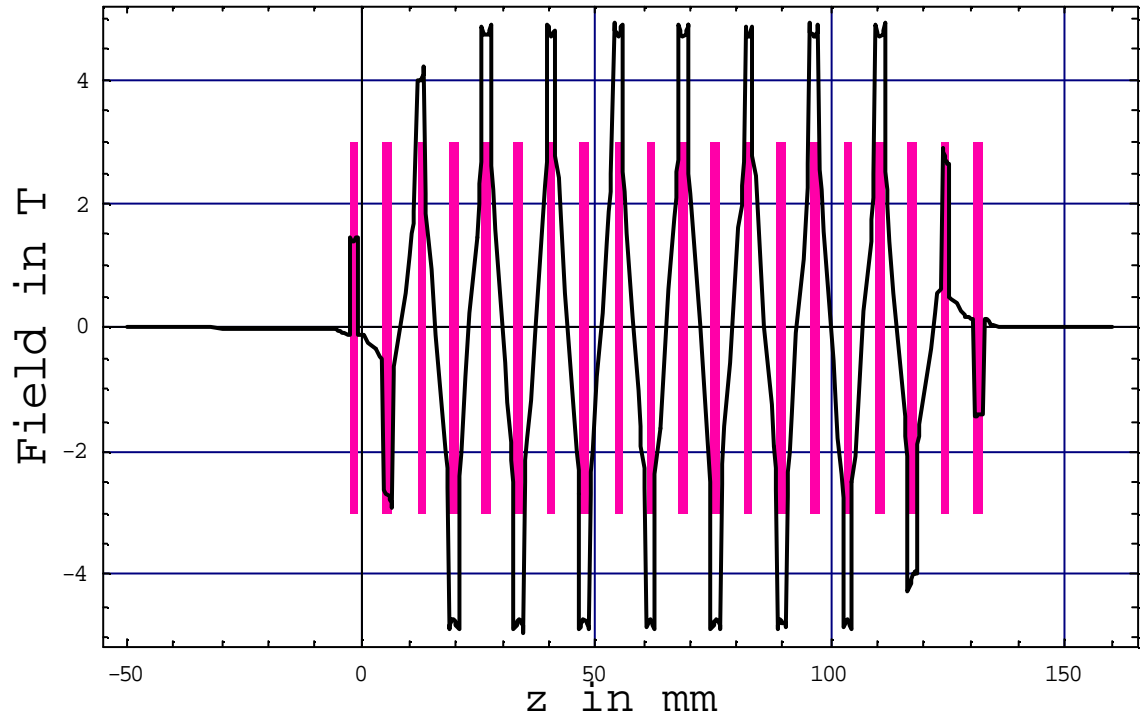

Fig. 8